\documentclass[aps,twocolumn,prl,amsfonts,showpacs,superscriptaddress,longbibliography]{revtex4-2}
\DeclareUnicodeCharacter{0393}{$\Gamma$}
\usepackage{verbatim,epsfig,amsmath,amssymb,bm,epsf,graphicx,psfrag,bbold,amsthm,amsfonts}
\usepackage[bottom]{footmisc}
\usepackage{hyperref}
\usepackage{cleveref} 
\usepackage{enumitem}
\usepackage{framed}
\usepackage{mathrsfs}
\usepackage{esint}
\setlist{nosep}
\usepackage{placeins}
\usepackage[all]{xy}
\usepackage{color}
\usepackage[utf8]{inputenc}
\usepackage{float}
\usepackage{natbib}
\usepackage{tikz-cd}
\usepackage{verbatim}
\usepackage{leftidx}
\usepackage{physics}
\usepackage{ulem}

\usepackage{pifont}
\usepackage{braket}
\usepackage{booktabs}

\newcommand{\psitheta}{\psi_{\scriptscriptstyle\theta}}

\newcommand{\kb}{{\bm k}}
\newcommand{\rb}{{\bm r}}
\newcommand{\qb}{{\bm q}}

\newcommand{\kk}{\bm k}

\newcommand{\aB}{a_{\scriptscriptstyle B}}

\begin{document}

\title{Accurate Self-Attention Wavefunctions at Large Scale}

\author{Filippo Gaggioli}
\affiliation{Department of Physics, Massachusetts Institute of Technology, Cambridge, MA-02139, USA}

\author{Sam Azadi}
\affiliation{Department of Physics and Astronomy, University of Manchester, Oxford Road, Manchester M13 9PL, UK}
\affiliation{Department of Physics, Clarendon Laboratory, University of Oxford, Parks Road, Oxford OX1 3PU, UK}

\author{Liang Fu}
\affiliation{Department of Physics, Massachusetts Institute of Technology, Cambridge, MA-02139, USA}

\date{\today}

\begin{abstract}
    
\end{abstract}

\begin{abstract}
    Self-attention neural networks provide powerful variational wavefunctions that surpass the expressivity of traditional variational ansätze. This expressivity, however, comes with increased computational complexity, raising a pressing question about scalability---can such wavefunctions retain their accuracy at large system sizes? We apply self-attention wavefunctions to the two-dimensional homogeneous electron gas for up to $N_e=169$ particles, obtaining energies systematically lower than state-of-the-art DMC.
    Direct access to the ground state wavefunction further lets us recover the full collective-mode dispersion of the liquid phase, from the small-$q$ plasmon branch to a roton-like minimum near $q\approx 2k_F$. Observables at $N_e=91$ and $N_e=169$ are in near-perfect agreement, indicating convergence to the thermodynamic limit.
\end{abstract}

\maketitle

Quantum mechanics provides the microscopic foundation of natural sciences. In condensed matter physics, it governs phenomena ranging from electronic band structure and magnetism to superconductivity, fractional quantum Hall states, and other strongly correlated phases. 
Quantum mechanics is also a quantitative predictive theory: in principle, the properties of atoms, molecules, and materials are entirely determined by the many-body Schrödinger equation for interacting nuclei and electrons.

In practice, however, computational access to this predictive power is severely limited by the exponential growth of the many-body Hilbert space, which represents the central obstacle faced by classical numerical methods for solving the ground state wavefunction in quantum many-body systems. In the study of correlated lattice systems, wavefunction-based approaches such as 
tensor-network methods \cite{White_1992} 
have enabled major progress on gapped groundstates, but struggle on gapless systems such as metals. 

Traditional variational Monte Carlo (VMC) techniques \cite{McMillan_1964, Ceperley_1976} represent a useful alternative for continuum fermionic systems \cite{Ceperley_1980, Foulkes_2001}.
In this approach, physically motivated trial states, typically built from Slater determinants augmented by Jastrow and backflow correlations \cite{Drummond_2004, LopezRios_2006}, can deliver accurate variational wavefunctions with a relatively compact set of parameters. These optimized trial states also play a central role as inputs to fixed-node diffusion Monte Carlo (DMC) \cite{Ceperley_1986}, where their nodal structure controls the remaining approximation. However, the accuracy of traditional VMC and DMC is limited by the quality of judiciously chosen trial wavefunctions.



Machine-learning-based methods have opened a new route to more powerful many-body solvers \cite{Carleo_2017, Carrasquilla_2017, Luo_2019, Pfau_2020}. Neural quantum states replace fixed analytic forms with trainable wavefunctions, allowing correlations to be learned directly through variational optimization. Among these approaches, self-attention architectures \cite{vonGlehn_2023, Geier_2025} are especially appealing because they can represent highly nonlocal and particle-dependent correlations without imposing a hand-crafted orbital ansatz. Rather than starting from a prescribed set of single-particle orbitals and modifying them through physically motivated transformations, self-attention wavefunctions can learn 
many-body correlation patterns purely from energy minimization.  
As a result, self-attention wavefunctions have proven capable of capturing different phases of matter within a unified ansatz \cite{Cassella_2023, Teng_2025, Qian_2025, Nazaryan_2025, Li_2025, Fadon_2025, Abouelkomsan_2026, Fu2026, Zhu_2026, Nazaryan_2026_qernel}, and even predicting new physics from first principles \cite{Gaggioli_2025}. 

The all-to-all character of the attention mechanism, which is central to its expressive power, comes, however, at the expense of a larger number of variational parameters and higher computational cost. This fundamental tension between expressivity and scalability raises a central question: \textit{can self-attention wavefunctions reach state-of-the-art accuracy at large system sizes, while retaining their greater expressivity?}

In this work, we address the question by applying self-attention wavefunctions to the two-dimensional homogeneous electron gas, for system sizes as large as $N_e=169$ particles. To demonstrate the accuracy of our self-attention wavefunctions, we compare variational energies with state-of-the-art DMC results and demonstrate systematically lower energies over a range of system sizes and interaction strengths. 

Our neural wavefunction approach gives full access to the ground-state, enabling the direct evaluation of many-body observables beyond the energy. We illustrate this by computing the momentum occupation and the static structure factor $S(q)$. From the structure factor, using the Feynman relation $\varepsilon(q)\sim q^2 / S(q)$, we extract the collective mode dispersion of the HEG, recovering the full excitation branch of the liquid phase: the plasmon dispersion at small momenta $q\ll k_F$  and a roton-like minimum at $q\approx 2 k_F$. Comparing results for $N_e=91$ and $N_e=169$ particles, we find that the extracted dispersions agree almost perfectly, indicating that our variational wavefunction remains accurate close to the thermodynamic limit.

\begin{figure}
    \centering    \includegraphics[width=0.9\linewidth]{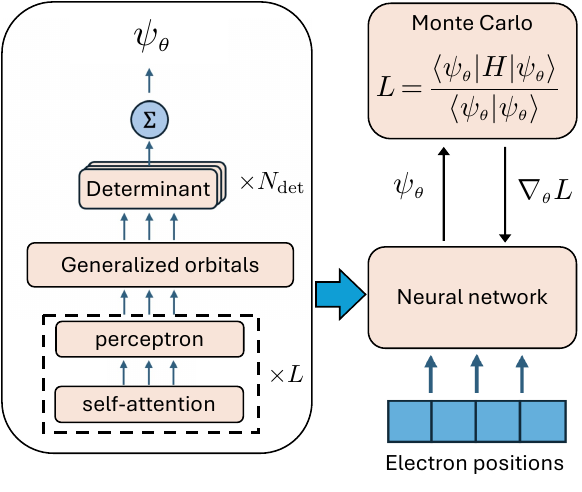}
    \caption{\textbf{Self-attention neural network and VMC:} Illustration of our self-attention architecture (left), and its role inside the VMC algorithm (right). 
    }
    \label{fig:architecture}
\end{figure}

\textit{Homogeneous electron gas ---}
We consider a system of spin-polarized electrons in two spatial dimensions, described in standard units by the Hamiltonian
\begin{equation}\label{eq:Hamiltonian}
    H = -\frac{\hbar^2}{2m}\sum_i\nabla_i^2
    + \frac{e^2}{4\pi\epsilon}\sum_{j<i}\frac{1}{|\mathbf r_i-\mathbf r_j|}.
\end{equation}
The interaction strength defines the effective Bohr radius
$\aB=4\pi\epsilon\hbar^2/me^2$, which sets the natural length scale of the problem. The dimensionless density parameter
$r_s = d/\aB$,
with
$d = \sqrt{1/\pi n}$ the average interparticle distance and $n$ the two-dimensional electron density, controls the ratio between interaction and kinetic energy. Small $r_s$ corresponds to the high-density metallic regime, while large $r_s$ favors crystallization. The best available estimates place the transition between the Fermi-liquid and Wigner-crystal phases near $r_s \approx 35$ \cite{Smith_2024,Azadi_2024}.

The competing liquid and crystalline ground states can be characterized through complementary many-body observables. The momentum distribution $n(\mathbf k)$ measures the occupation of plane-wave orbitals $e^{i\mathbf k\cdot\mathbf r}$ \cite{supplementary} -- in the metallic phase, it exhibits a sharp drop at the Fermi wavevector
$k_F = \sqrt{4\pi n}$.
Spatial order is instead probed by the static structure factor $S(\qb) = \langle \rho_\qb \rho_{-\qb}\rangle /N_e $, with $\rho_\qb =  \sum_i e^{i\mathbf q\cdot\mathbf r_i}$ the density operator at momentum $\qb$.
%
In the crystalline phase, $S(\mathbf q)$ develops sharp Bragg peaks at the reciprocal-lattice vectors of the triangular Wigner crystal, located at momenta
$|\mathbf q| = \sqrt{8\pi n/\sqrt{3}}$.

In addition, the structure factor
%
%
encodes information about collective density fluctuations above the ground state. As first pointed out by Feynman in the context of superfluid $^4$He \cite{Feynman_1954}, a variational estimate for the energy of a density excitation with wavevector $\mathbf q$ is given by
\begin{equation}\label{eq:Feynman}
    \varepsilon(\mathbf q)
    \lesssim
    \frac{\hbar^2 q^2}{2m S(\mathbf q)} .
\end{equation}
Large values of $S(\mathbf q)$ therefore signal soft density modes. This relation is especially useful at small wavevectors $q\ll k_F$, where the collective-mode dispersion reflects the nature of the ground state. For the Coulomb-interacting electron gas, the long-wavelength density mode is the plasmon, whose dispersion scales as $\varepsilon(q)\propto \sqrt{q}$ in two dimensions.

A systematic analysis of these observables across different values of $r_s$ is presented below, after introducing the self-attention neural-network wavefunction used in our variational Monte Carlo calculations.

\begin{table}[t]
\centering
\caption{Energy comparison (inference).}
\label{tab:energies}
\begin{tabular}{c c c c}
\hline
System & Method & $E$ [Ha/el] & $E_\text{corr}$ [Ha/el] \\
\hline
$N_e=91$     & SJB-VMC & $-0.0318792(2)$ & $-0.00456940$ \\
$r_s=30$     & SJB-DMC & $-0.0319370(1)$ & $-0.00462720$ \\
              & \textbf{NN-VMC} & $\mathbf{-0.03193763(3)}$ & $\mathbf{-0.0046278}$ \\
\hline
$N_e=91$     & SJB-VMC & $-0.0243832(1)$ & $-0.00369277$ \\
$r_s=40$     & SJB-DMC & $-0.0244377(2)$ & $-0.00374727$ \\
              & \textbf{NN-VMC} & $\mathbf{-0.02443814(2)}$ & $\mathbf{-0.0037477}$ \\
\hline
$N_e=169$    & SJB-VMC & $-0.0318548(2)$ & $-0.00464301$ \\
$r_s=30$     & SJB-DMC & $-0.0319197(1)$ & $-0.00470791$ \\
              & \textbf{NN-VMC} & $\mathbf{-0.03192077(3)}$ & $\mathbf{-0.0047090}$ \\
\hline
\end{tabular}
\end{table}

\begin{figure*}
    \centering
    \includegraphics[width=\textwidth]{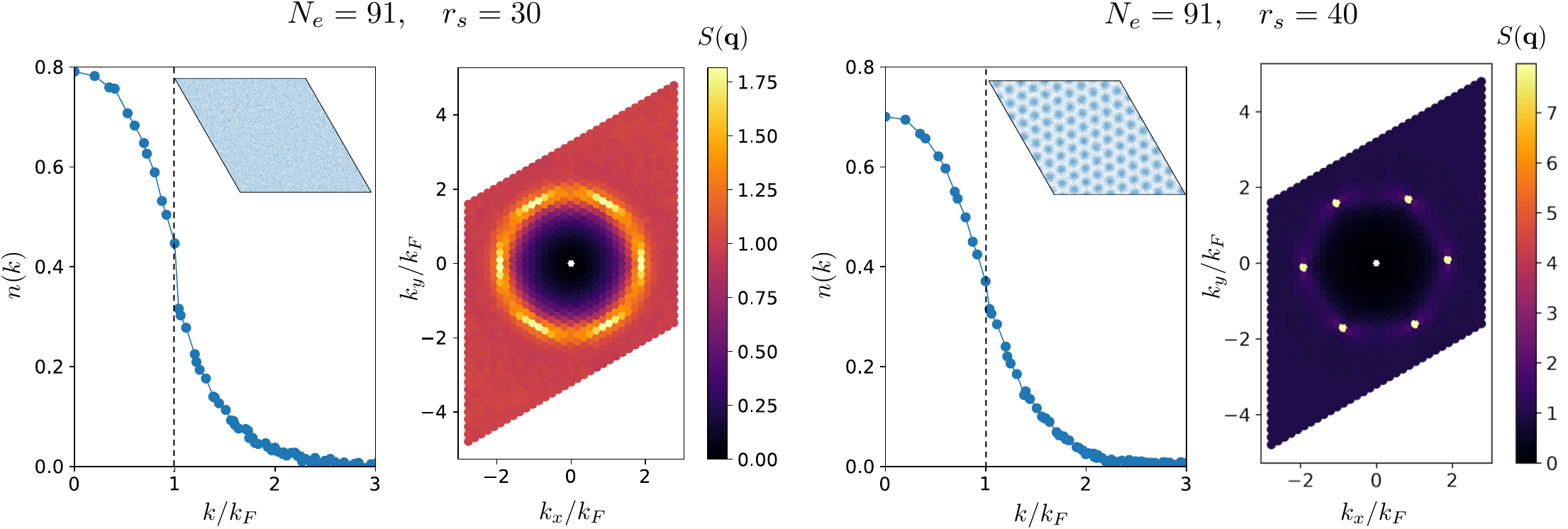}
    \caption{\textbf{Momentum space occupation and structure factor.} Left panels: for $r_s = 30$, the system displays metallic behavior, as signaled by the sharp Fermi surface discontinuity in the momentum space occupations $n(k)$. Right panels: for $r_s = 40$, the electrons form a crystalline state (see the density snapshot in the inset). Correspondingly, prominent Bragg peaks appear in the structure factor $S(\qb)$.  }
    \label{fig:main}
\end{figure*}

\textit{Self-attention wavefunction ---}
Our neural-network ansatz \cite{vonGlehn_2023,Geier_2025} is illustrated in Fig.\ \ref{fig:architecture}. It consists of a stack of self-attention and perceptron layers, repeated $L$ times, which takes the electron positions $\rb_j$ as input and outputs a set of vectors that, after a simple projection, define generalized single-particle orbitals $\phi^{(k)}_i\left(\rb_j,\{\rb_{\neq j}\}\right)$. Here, the notation emphasizes that the orbital associated with particle $j$ depends not only on its own position $\rb_j$, but also on the full many-body configuration of the remaining electrons $\{\rb_{\neq j}\}$. These generalized orbitals are then combined into $N_{\rm det}$ Slater determinants, whose sum defines the antisymmetric fermionic neural wavefunction
\begin{equation}\label{eq:self-attention_ansatz}
    \psitheta(\rb_1,\cdots,\rb_N)
    =
    \sum_{k=1}^{N_{\rm det}}
    \det\left[
    \phi_i^{(k)}
    \left(\rb_j,\{\rb_{\neq j}\}\right)
    \right],
\end{equation}
parametrized by the network weights $\theta$. 

Notably, the self-attention ansatz \eqref{eq:self-attention_ansatz} does not rely on a prescribed 
set of orbitals
, as opposed to message-passing neural network architectures that implements the backflow transformation on plane-wave orbitals \cite{Pescia_2024, Smith_2024, Valenti_2025, Linteau_2026}. Instead, it generates generalized orbitals directly from the many-body electron configuration. Moreover, the way in which each particle is effectively ``dressed'' by the surrounding electrons is not constrained to be the same for all orbitals, as 
in standard backflow coordinate-transformations. In this sense, the self-attention architecture both avoids the bias associated with a prescribed choice of single-particle orbitals and provides a more flexible representation of many-body correlations.

The all-to-all character of the attention mechanism \cite{Vaswani_2017} presents a challenge for scaling the self-attention ansatz to large system sizes. On physical grounds, correlations decay with spatial separation, implying that, in its vanilla implementation, the transformer architecture must invest a large amount of ``attention'' to learn how distant particles are \textit{not} strongly influencing each other. 

To overcome this limitation, we introduce an exponential damping factor that biases the attention mechanism according to the interparticle distance $r_{ij}$. In terms of the usual  ``key'' and ``query'' matrices $K,Q$, 
the pre-softmax attention logit of each particle pair is then defined as
\begin{equation}\label{eq:attention_bias}
    \ell_{ij} = \frac{Q_i\cdot K_j}{\sqrt{d_{\scriptscriptstyle K}}}  -\frac{r_{ij}}{\lambda},
\end{equation}
where $ i,j = 1, \cdots, N_e$ are the particle indices and $d_{\scriptscriptstyle K}$ is the dimension of the key and query.
The trainable lengthscale  $\lambda$ is initialized from the average interparticle distance $\sqrt{1 / \pi n}$, without further problem-specific tuning. 
Our approach is inspired by  Refs.~\cite{Zhang_2025, Viteritti_2026}, which introduced ``spatial attention'' based on the fixed distances between lattice sites for studying spin systems. 
In contrast, our attention bias is associated with pairs of particle positions that constitute the input to our first-quantized wavefunction, which is applicable to both continuum and lattice systems \cite{Fu2026}. For this reason, we refer to Eq.~\eqref{eq:attention_bias} as \textit{``spatially weighted particle attention''}.


We finally remark that, since the attention operation is applied repeatedly across several layers, particle information can progressively propagate beyond the scale set by the damping length $\lambda$. In this way, correlations can build up over increasingly large distances through the depth of the network. The locality bias in Eq.~\eqref{eq:attention_bias} therefore does not prevent the ansatz \eqref{eq:self-attention_ansatz} from capturing correlations arising from long-range interactions, as demonstrated by the results in the next section.

\textit{Results ---} We now solve the Hamiltonian \eqref{eq:Hamiltonian} in a triangular simulation cell, using periodic boundary conditions at the $\Gamma$-point. The long-range Coulomb interaction is treated through Ewald summation \cite{CASINO2019manual}. We focus on spin-polarized systems with $N_e=91$ and $N_e=169$ electrons, and consider values of $r_s=30,40$ close to the liquid--crystal transition, where correlation effects are strongest and obtaining accurate wavefunctions is most challenging.

To benchmark the accuracy of our self-attention wavefunction, we compare the variational energies against state-of-the-art diffusion Monte Carlo results. The NN-VMC calculations were performed using the self-attention architecture in Fig.~\ref{fig:architecture}, with $\approx 800\,$K parameters. The training lasted about $100\,$K steps, and ran on two or four H$200$ GPUs. 
The DMC benchmarks (SJB-DMC) were obtained within the fixed-node approximation, using optimized Slater-Jastrow-backflow wavefunctions (SJB-VMC) to impose the nodal constraint, and were extrapolated to the zero-time-step limit to remove the leading time-step bias. We refer to the SM \cite{supplementary} and Ref.~\cite{Azadi_2024} for further details on the DMC methodology.

The results are summarized in Tab.\ \ref{tab:energies}. For interaction strengths considered, our NN-VMC energies are \textit{systematically lower and more precise} than the corresponding fixed-node DMC estimates, even for the largest system size of $N_e=169$.
Expressing energies in units of the correlation energy $E_\text{corr}$ (defined with respect to the Hartree-Fock energies at the same system size), we quantify the NN-VMC margin over the traditional SJB-VMC benchmark as $1.263\%$ and $1.466\%$ for $N_e = 91$ with $r_s = 30,40$, and $1.401\%$ for the largest system size with $N_e = 169$ electrons. Remarkably, the accuracy of the self-attention wavefunction does not deteriorate upon increasing the system size.

On top of its accuracy, the explicit nature of the NN-VMC wavefunction allows us to compute ground-state observables beyond the energy. Fig.\ \ref{fig:main} shows the momentum occupation $n(\mathbf k)$ and the static structure factor $S(\mathbf q)$ for $N_e=91$ electrons at $r_s=30$ (liquid) and $r_s=40$ (crystal) -- the corresponding densities are shown in the smaller insets. The momentum distribution displays a sharp Fermi surface in the liquid phase as evidenced by the discontinuous jump in $n(\mathbf k)$ across $k_F$, accompanied by a finite interaction-induced tail beyond $k_F$. 
The crystalline phase, on the other hand, displays a smooth and wider occupation curve, without visible Fermi surface discontinuity around $k_F$.

The structure factor provides complementary information about spatial density correlations. At $r_s=30$, $S(\mathbf q)$ shows enhanced weight on an approximately circular shell around $2k_F$, but no sharp crystalline peaks. This is consistent with a strongly correlated liquid close to crystallization. At $r_s=40$, instead, the structure factor develops pronounced peaks at six symmetry-related momenta, signaling the emergence of triangular Wigner-crystal order. This is also visible directly from the real-space density snapshots shown in the insets of Fig.\ \ref{fig:main}: while the $r_s=30$ state remains essentially homogeneous, the $r_s=40$ state displays a clear triangular modulation.

\begin{figure}
    \centering\includegraphics[width=0.9\linewidth]{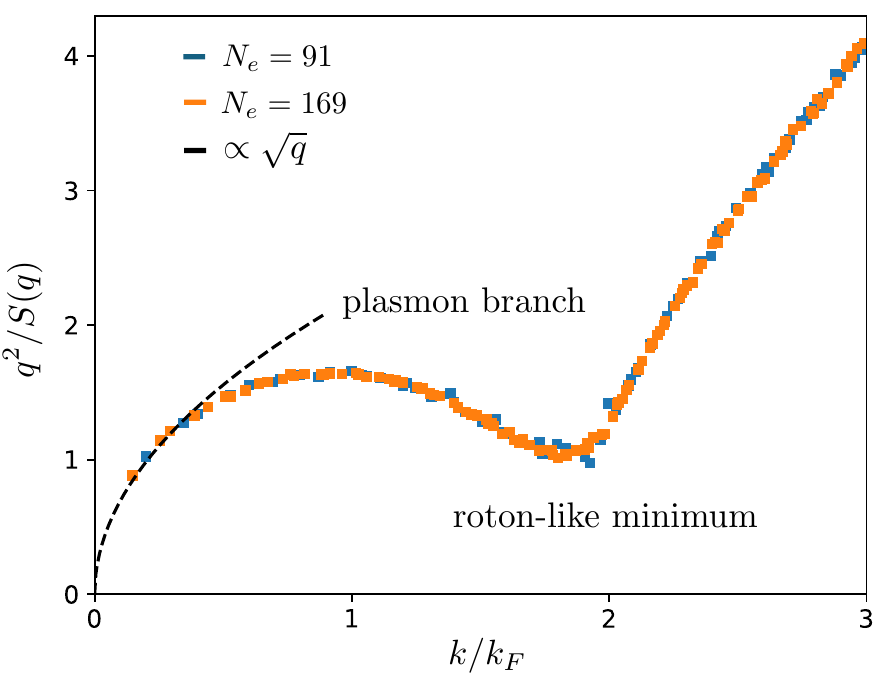}
    \caption{\textbf{Collective modes ($r_s = 30$):} the energy $q^2 / S(q)$ provides a variational upper bound on the dispersion of the collective modes  for the HEG ground state. The plasmon branch $\propto \sqrt{q}$ at small $q$ evolves into a roton-like minimum around $2k_F$ associated to the incipient crystalline order. The quantitative agreement across system sizes certifies accuracy in approaching the thermodynamic limit. }
    \label{fig:plasmon}
\end{figure}

Finally, we use the structure factor to extract information about collective density modes of the liquid phase using equation \eqref{eq:Feynman}.
The resulting variational upper bound on the excitation dispersion $q^2/S(q)$, obtained by taking the angular average of $S(\qb)$ at $r_s = 30$, is shown in Fig.\ \ref{fig:plasmon} for $N_e=91$ and $N_e=169$. At small wavevectors, the dispersion follows the expected two-dimensional plasmon scaling $\varepsilon(q)\propto \sqrt{q}$, confirming that the neural wavefunction \eqref{eq:self-attention_ansatz} correctly capture the long-wavelength Coulomb correlations of the HEG. At larger momenta, the dispersion develops a clear roton-like minimum close to $2k_F$, reflecting the softening of density fluctuations at the wavevector associated with the incipient crystalline order.

Remarkably, the curves obtained for $N_e=91$ and $N_e=169$ are almost indistinguishable over the full range of momenta. This agreement shows that the collective-mode dispersion extracted from the self-attention wavefunction already converges to its thermodynamic-limit behavior at the system sizes considered here. Together with the variational energies below fixed-node DMC, these results demonstrate that self-attention NN-VMC provides not only accurate ground-state energies, but also reliable access to correlated observables and collective excitations in the continuum homogeneous electron gas at very large system sizes.

\textit{Discussion ---} The availability of expressive variational wavefunctions is an essential ingredient for discovering new physics with computational tools. At the same time, resolving the delicate interplay of energy scales in many-body systems requires quantitative control at the highest level of numerical accuracy. Our results show that self-attention architectures can simultaneously meet both requirements up to large system sizes.

Success on the homogeneous electron gas problem – a cornerstone of many-body condensed matter physics – opens the door to using self-attention architectures to explore new physics in semiconductor systems and beyond. 
More broadly, access to large system sizes enables the study of increasingly realistic electronic systems, with potential applications ranging from materials discovery to quantum technologies.

\textit{Acknowledgments --} We thank Ahmed Abouelkomsan for insightful comments.  This work was supported by the National Science Foundation (NSF) Convergence Accelerator Award No. 2235945. 
We acknowledge IAIFI (the NSF Cooperative Agreement PHY-2019786) and the MIT Engaging cluster for providing the computational resources that have contributed to the NN-VMC results reported in this paper.   
L.F. was supported by a Simons Investigator Award from the Simons Foundation. S.A. acknowledges the support of the Leverhulme Trust under the grant agreement RPG-2023-253.

\bibliography{biblio.bib}

\onecolumngrid
\newpage
\makeatletter

\begin{center}
\textbf{\large Supplementary materials for:\\ ``{Accurate Self-Attention Wavefunctions at Large Scale} ''}
\\[10pt]
Filippo Gaggioli$^{1}$, 
Sam Azadi$^{2}$,
Liang Fu$^{1}$\\
\textit{$^1$Department of Physics, Massachusetts Institute of Technology, Cambridge, MA-02139, USA\\}
\textit{$^2$Department of Physics and Astronomy, University of Manchester, Oxford Road, Manchester M13 9PL, UK}\\
\textit{$^2$Department of Physics, Clarendon Laboratory, University of Oxford, Parks Road, Oxford OX1 3PU, UK}
\end{center}
\vspace{20pt}

\setcounter{figure}{0}
\setcounter{section}{0}
\setcounter{equation}{0}

\renewcommand{\thefigure}{S\@arabic\c@figure}
\makeatother

\section{Momentum-space occupation}

In this section, we give the recipe used to evaluate the momentum-space occupations $n(\kb)$ displayed in Fig.~\ref{fig:main}. These are read off from the diagonal of the one-body reduced density matrix (1RDM). 

We work in a single particle basis of plane-waves,
\begin{align}
    \Phi_{\kb}(\rb) = \frac{1}{\sqrt{A}}e^{i \kb \cdot \rb},
\end{align}
where $A = | \mathbf{L}_1\times \mathbf{L}_2|$ is the area of the supercell with arms $\mathbf{L}_1$ and $ \mathbf{L}_2$. Periodic boundary conditions restrict the momenta to the supercell reciprocal grid
\begin{equation}
  \kk_{mn} \;=\; m\,\mathbf{b}_1 + n\,\mathbf{b}_2,
  \qquad
  B = (\mathbf{b}_1, \mathbf{b}_2) = 2\pi \left(L^{-1}\right)^{\!T},
  \qquad
  \mathbf{b}_i \cdot \mathbf{L}_j = 2\pi\,\delta_{ij},
\end{equation}
with $m, n \in \mathbb{Z}$ and $L = (\mathbf{L}_1, \mathbf{L}_2)$ the supercell lattice matrix. 

In this basis, the 1RDM has elements
\begin{align}
    \rho_1(\kb; \kb') &= \sum_{i=1}^{N} \int \text{d}^2\rb_{i} \,\text{d}^2\rb'_i \left( \prod_{j \neq i } \text{d}^2\rb_j \right)\ \Psi(\cdots,\ \rb_i,\ \cdots) \Psi^{*}(\cdots,\  \rb_i',\ \cdots) \Phi^{*}_{\kb}(\rb_i) \Phi_{\kb'}(\rb_i')\nonumber\\
    &= \sum_{i=1}^{N} \int \text{d}^2\rb_i \left( \prod_{j \neq i } \text{d}^2\rb_j \right) \,\  |\Psi(\cdots,\ \rb_i,\ \cdots)|^2 \int \text{d}'\rb_i  \frac{\Psi^{*}(\cdots,\ \rb_i',\ \cdots)}{\Psi^{*}(\cdots,\ \rb_i,\ \cdots)} \Phi^{*}_{\kb}(\rb_i) \Phi_{\kb'}(\rb_i').
\end{align}
On the second line, the elements are expressed in a form that can be evaluated by Monte Carlo sampling a batch of configurations $\mathcal{B}$ from the probability distribution $|\Psi|^2$, yielding
\begin{align}
     \rho_1(\kb; \kb') &\approx \sum_{i=1}^{N_e}\sum_{\{\rb_i\} \in \mathcal{B}} \int \text{d}'\rb_i \,  \frac{\Psi^{*}(\cdots,\ \rb_i',\ \cdots)}{\Psi^{*}(\cdots,\ \rb_i,\ \cdots)} \,\Phi^{*}_{\kb}(\rb_i) \Phi_{\kb'}(\rb_i').
\end{align}
The two-dimensional momentum-space occupation is then the corresponding diagonal element, obtained by setting $\kb' = \kb$,
\begin{align}
    n(\kb) \equiv \rho_1(\kb; \kb) \approx \sum_{i=1}^{N_e}\sum_{\{\rb_i\} \in \mathcal{B}} \int \text{d}'\rb_i  \, \frac{\Psi^{*}(\cdots,\ \rb_i',\ \cdots)}{\Psi^{*}(\cdots,\ \rb_i,\ \cdots)} \Phi^{*}_{\kb}(\rb_i) \Phi_{\kb}(\rb_i'),
\end{align}
and is normalized such that $\sum_{\kb} n(\kb) = N_e$.

\section{Diffusion Monte Carlo}
\begin{figure}[!htb]
    \centering
    \includegraphics[width=0.5\linewidth]{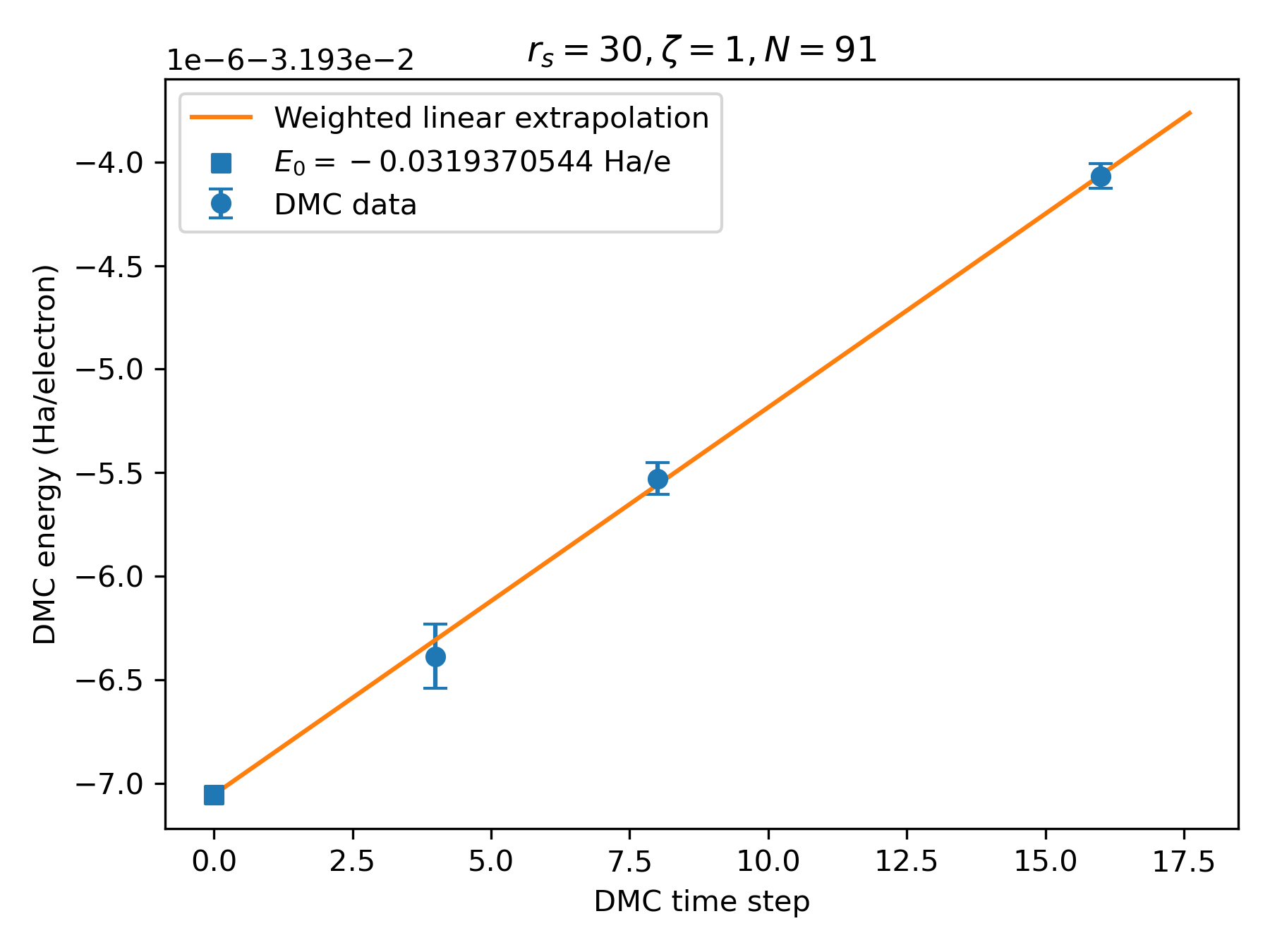}
    \includegraphics[width=0.5\linewidth]{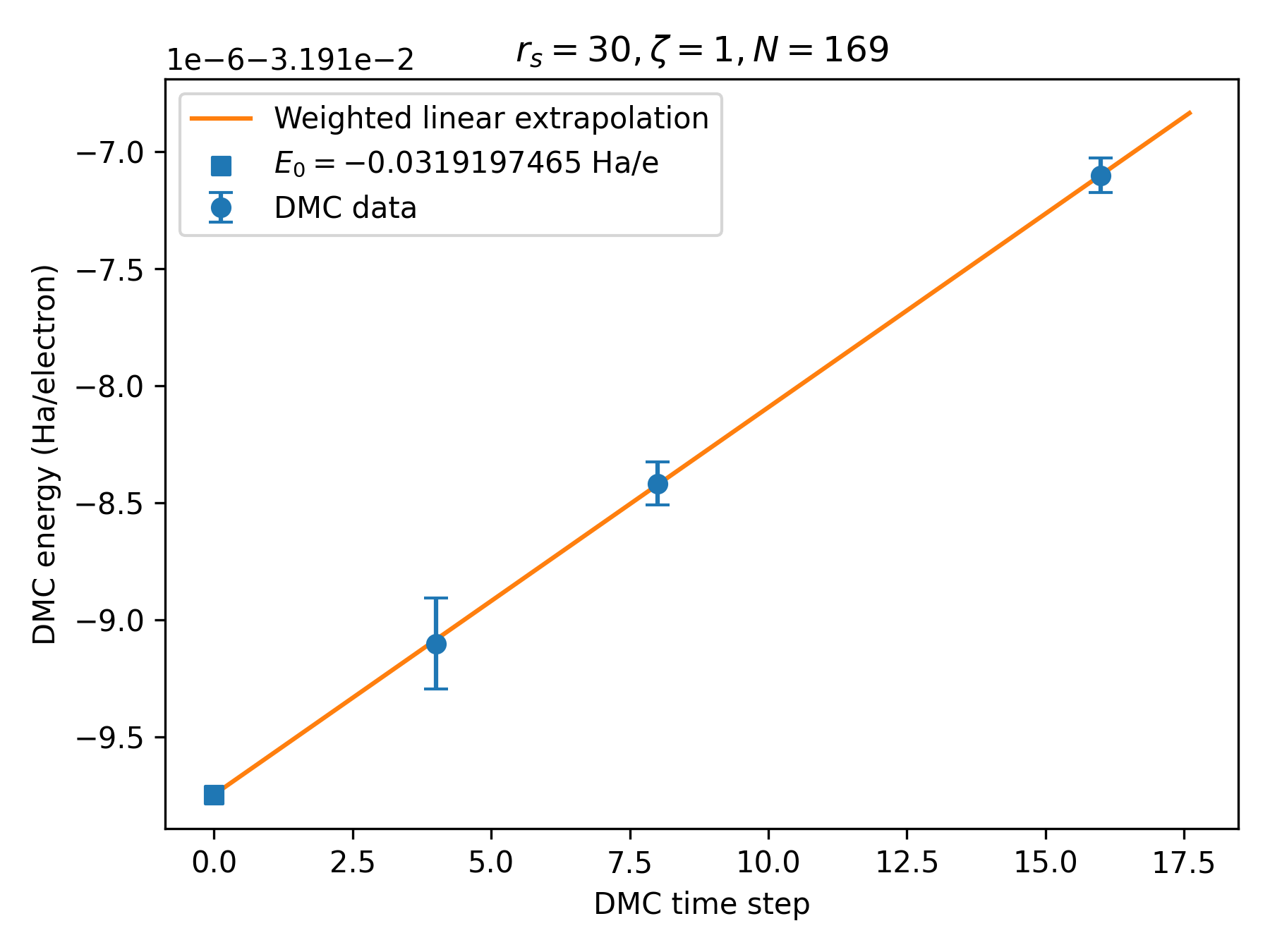}
    \includegraphics[width=0.5\linewidth]{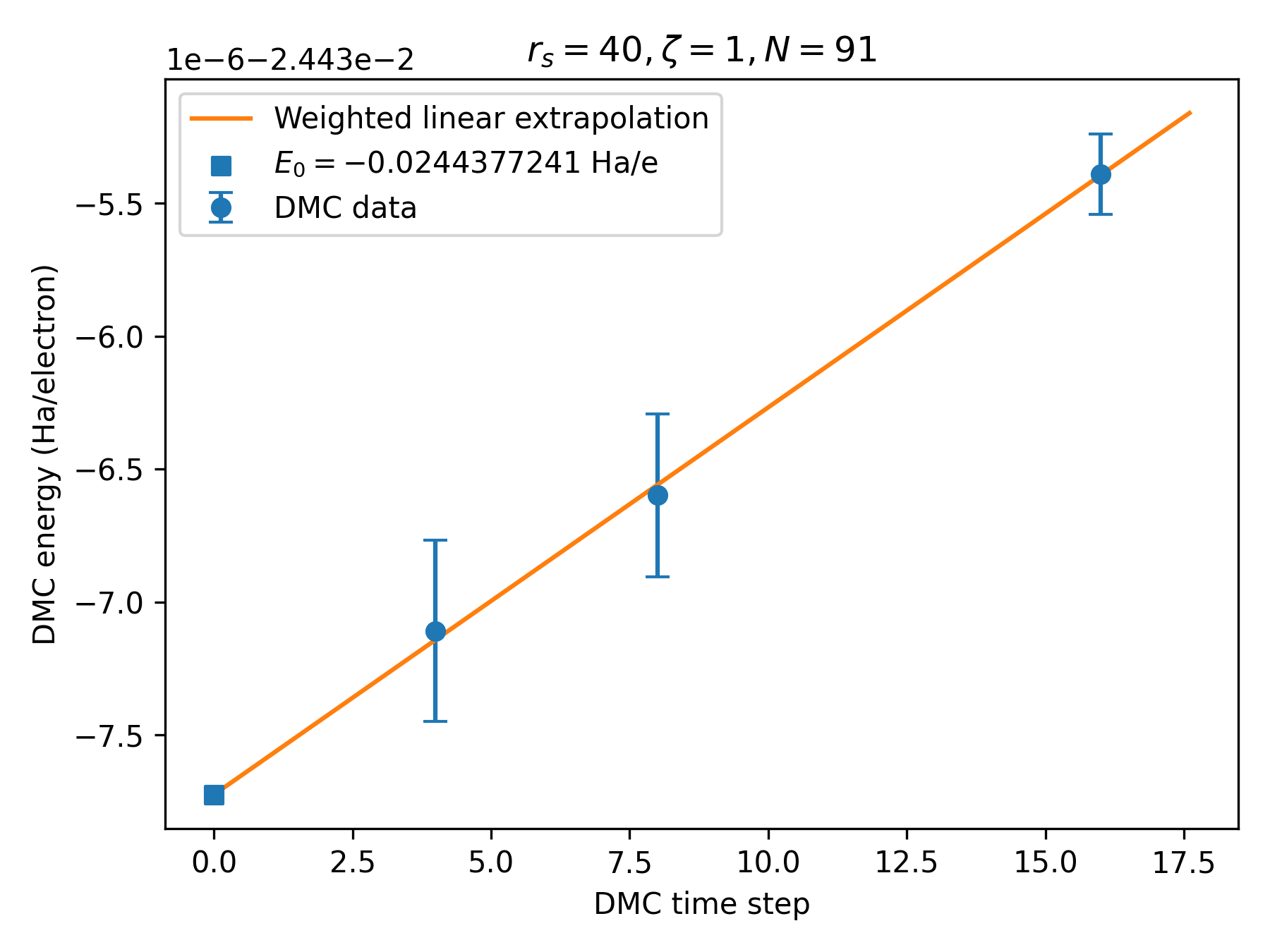}
    
    \caption{Linear extrapolation of the DMC energies as a function of time step to obtain the zero-time-step limit.}
    \label{fig:dt_extrapol}
\end{figure}

Details of our variational and diffusion Monte Carlo simulations are provided in Ref.~\cite{Azadi_2024}. We employed the same Slater–Jastrow–backflow (SJB) trial wave function and the same simulation-cell symmetry as used previously for the liquid phase. The only difference is that the VMC and DMC energies reported here were obtained at the $\Gamma$-point. To remain consistent with the neural-network setup, we did not apply twist-averaged boundary conditions. The DMC energies were extrapolated to the zero-time-step limit (Fig.~\ref{fig:dt_extrapol}).

\end{document}